\documentclass[12pt,preprint]{aastex}

\usepackage{natbib}
\bibpunct{(}{)}{,}{a}{}{,}
\usepackage{array}
\usepackage{amsmath}

\newcommand{\teff}{T_\mathrm{eff}}
\newcommand{\feh}{\mathrm{[Fe/H]}}
\newcommand{\logg}{\log g}
\newcommand{\msun}{M_\odot}
\newcommand{\as}{asteroseismology }

\shorttitle{Verification of self-consistent asteroseismic stellar parameters}

\begin{document}
\title{VERIFYING ASTEROSEISMICALLY DETERMINED PARAMETERS OF {\it KEPLER} STARS USING HIPPARCOS PARALLAXES: SELF-CONSISTENT STELLAR PROPERTIES AND DISTANCES}

\author{V.~Silva~Aguirre\altaffilmark{1,2,3}, L.~Casagrande\altaffilmark{4}, S.~Basu\altaffilmark{5}, T.~L.~Campante\altaffilmark{6,7}, W.~J.~Chaplin\altaffilmark{7,1,3}, D.~Huber\altaffilmark{8}, A.~Miglio\altaffilmark{7}, A.~M.~Serenelli\altaffilmark{9,3},  J.~Ballot\altaffilmark{10,11}, T.~R.~Bedding\altaffilmark{12,1}, J.~Christensen-Dalsgaard\altaffilmark{1,3}, O.~L.~Creevey\altaffilmark{13}, Y.~Elsworth\altaffilmark{7}, R.~A.~Garc\'ia\altaffilmark{14,3}, R.~L.~Gilliland\altaffilmark{15}, S.~Hekker\altaffilmark{16,7}, H.~Kjeldsen\altaffilmark{1}, S.~Mathur\altaffilmark{17,3}, T.~S.~Metcalfe\altaffilmark{18}, M.~J.~P.~F.~G.~Monteiro\altaffilmark{6}, B.~Mosser\altaffilmark{19}, M.~H.~Pinsonneault\altaffilmark{20,3}, D.~Stello\altaffilmark{12}, A.~Weiss\altaffilmark{2}, P.~Tenenbaum\altaffilmark{21}, J.~D.~Twicken\altaffilmark{21}, and K.~Uddin\altaffilmark{22}}
\altaffiltext{1}{Stellar Astrophysics Centre, Department of Physics and Astronomy, Aarhus University, Ny Munkegade 120, DK-8000 Aarhus C, Denmark}
\altaffiltext{2}{Max Planck Institute for Astrophysics, Karl-Schwarzschild-Str. 1, D-85748 Garching bei M\"{u}nchen, Germany}
\altaffiltext{3}{Kavli Institute for Theoretical Physics, Santa Barbara, CA 93106, USA}
\altaffiltext{4}{Research School of Astronomy and Astrophysics, Mount Stromlo Observatory, The Australian National University, ACT 2611, Australia }
\altaffiltext{5}{Department of Astronomy, Yale University, P.O. Box 208101, New Haven, CT 06520-8101, USA}
\altaffiltext{6}{Centro de Astrof\'{\i}sica and Faculdade de Ci\^encias, Universidade do Porto, Rua das Estrelas, 4150-762 Porto, Portugal}
\altaffiltext{7}{School of Physics and Astronomy, University of Birmingham, Birmingham B15 2TT, UK}
\altaffiltext{8}{NASA Ames Research Center, Moffett Field, CA 94035, USA}
\altaffiltext{9}{Instituto de Ciencias del Espacio (CSIC-IEEC), Facultad de  Ci\`encies, Campus UAB, E-08193 Bellaterra, Spain}
\altaffiltext{10}{CNRS, Institut de Recherche en Astrophysique et Plan\'etologie, 14 avenue Edouard Belin, F-31400 Toulouse, France}
\altaffiltext{11}{Universit\'e de Toulouse, UPS-OMP, IRAP, F-31400 Toulouse, France}
\altaffiltext{12}{Sydney Institute for Astronomy, School of Physics, University of Sydney, NSW 2006, Australia}
\altaffiltext{13}{Laboratoire Lagrange, UMR 7293, Universit\'e de Nice Sophia-Antipolis, CNRS, Observatoire de la C\^ote dÕAzur, BP 4229, 06304 Nice Cedex 4, France}
\altaffiltext{14}{Laboratoire AIM, CEA/DSM-CNRS-Universit\'e Paris Diderot, IRFU/SAp, Centre de Saclay, F-91191 Gif-sur-Yvette Cedex, France}
\altaffiltext{15}{Center for Exoplanets and Habitable Worlds, The Pennsylvania State University, University Park, PA, USA}
\altaffiltext{16}{Astronomical Institute "Anton Pannekoek," University of Amsterdam, Science Park 904, 1098 XH Amsterdam, The Netherlands}
\altaffiltext{17}{High Altitude Observatory, NCAR, P.O. Box 3000, Boulder, CO 80307, USA}
\altaffiltext{18}{Space Science Institute, Boulder, CO 80301, USA}
\altaffiltext{19}{LESIA, CNRS, Universit\'e Pierre et Marie Curie, Universit\'e Denis Diderot, Observatoire de Paris, F-92195 Meudon Cedex, France}
\altaffiltext{20}{Department of Astronomy, The Ohio State University, Columbus, OH 43210, USA}
\altaffiltext{21}{SETI Institute/NASA Ames Research Center, Moffett Field, CA 94035, USA}
\altaffiltext{22}{Orbital Sciences Corporation/NASA Ames Research Center, Moffett Field, CA 94035, USA}

\begin{abstract}
Accurately determining the properties of stars is of prime importance for characterizing stellar populations in our Galaxy. The field of asteroseismology has been thought to be particularly successful in such an endeavor for stars in different evolutionary stages. However, to fully exploit its potential, robust methods for estimating stellar parameters are required and independent verification of the results is mandatory. With this purpose, we present a new technique to obtain stellar properties by coupling asteroseismic analysis with the InfraRed Flux Method. By using two global seismic observables and multi-band photometry, the technique allows us to obtain masses, radii, effective temperatures, bolometric fluxes, and hence distances for field stars in a self-consistent manner. We apply our method to 22 solar-like oscillators in the {\it Kepler} short-cadence sample, that have accurate {\it Hipparcos} parallaxes. Our distance determinations agree to better than 5\%, while measurements of spectroscopic effective temperatures and interferometric radii also validate our results. We briefly discuss the potential of our technique for stellar population analysis and models of Galactic Chemical Evolution.
\end{abstract}

\section{Introduction}\label{intro}
Studying the structure and evolution of the Milky Way requires detailed knowledge of the properties of the stellar populations comprising it. In this respect, \as is a powerful tool to determine masses and radii of single stars to a high level of precision \citep[e.g.,][]{bm10,tk10,tm10}. The {\it CoRoT} \citep{ab06,em08} and \textit{Kepler} missions \citep{gill10,wb10,dk10} have provided data on stellar oscillations of exquisite quality for thousands of stars, encouraging us to carry out a complete stellar census of the observed populations.

From the thousands of light curves obtained by the space missions, two asteroseismic parameters can be readily extracted \citep[e.g.,][]{sh11a,dh11,wc11}. First, the power spectrum of solar-like oscillators is modulated in frequency by a Gaussian-like envelope, where the frequency of maximum power $\nu_{\rm{max}}$ scales approximately with the surface gravity and effective temperature. Second, the near-regular pattern of high overtones presents a dominant frequency spacing called the large frequency separation, $\Delta\nu$, which scales approximately with the square root of the mean stellar density. Applying scaling relations from solar values, these two asteroseismic observables may be used to estimate stellar properties of large numbers of solar-like oscillators, where individual frequencies are not available for all targets \citep[e.g.,][]{ds08,sb10,sh11a,dh11,vsa11a}.

To gain insight about the formation history and evolution of our Galaxy, characteristics of stellar populations distributed across it must be accurately known. The best studied sample of stars in the Milky Way is the solar neighborhood, where observations and analysis over many years have determined some of its key properties, such as orbits, kinematics, and metallicities \citep[e.g.,][]{be93,br03,bn04,fv07,fb08,lc11}. These data comprise the basic set of constraints for any model of chemical evolution of the Galaxy.

Models of Galactic Chemical Evolution are constructed under certain assumptions regarding the physical processes involved in the evolution of our Galaxy, and then calibrated against available observations. Those that reproduce them successfully are also used to predict other properties of the Galaxy, such as abundance gradients across the disk, gas infall episodes, and star formation rates \citep[e.g.,][]{bt80,cc97,lp98,sb09}. Thus, their predictive power for our galactic history and morphology critically depends on how well they can reproduce these observations, most of which come from the solar neighborhood sample. Of particular importance among these restrictions is the age--metallicity relation, constructed using stellar isochrones and determinations of element abundances \citep[e.g.,][]{be93,bn04}. The existence of an age--metallicity relation in the solar neighborhood is still a subject of debate \citep[see][and references therein]{sf01,bn04,kf12}, and accurate age determinations are of prime importance to shed some new light in this issue. However, the solar neighborhood sample used to constrain these models is only complete to distances of $\sim$50~pc \citep[e.g.,][]{bn04}, and accurate properties of stars further than $\sim$100~pc are difficult to measure, yet of crucial importance \citep[e.g.,][]{fb02,mst06,zi08}. To extend the sample used as a testbed for comparison, we need stellar parameters measured with high accuracy in different regions of the Galaxy.

Asteroseismology can help bridge this gap by providing accurate stellar properties, including distances, for field stars out to several hundred parsecs. These parameters, together with effective temperatures, metallicities, and kinematics, should make possible to study spatial gradients of stellar properties across the Galactic disk and provide insight into the formation process of our Galaxy \citep[see][]{am12a,jc12,am12c}. Moreover, robust age determinations obtained combining this information with evolutionary models will allow construction of the age-metallicity relation of the stellar populations observed by {\it CoRoT} and {\it Kepler}, and so provide tests outside the solar neighborhood of Galactic Chemical Evolution models.

The first comparison between asteroseismically determined parameters and predictions from Galactic Chemical Evolution models was made by \citet{am09} for a sample of {\it CoRoT} red giants. \citet{wc11} used the same technique to obtain masses and radii of \textit{Kepler} main-sequence and subgiant stars, and found a slight but statistically significant difference between the observed and synthetic mass distributions. Continuing this line of work, \citet{am12b} obtained distances to {\it CoRoT} and \textit{Kepler} red giants using bolometric corrections retrieved from the literature, while \citet{oc12} took a similar approach for five {\it Kepler} subgiant stars.

Considering the enormous potential of the results, it is important to verify the techniques applied in asteroseismic analysis. So far, empirical tests of the scaling relations have used heterogeneous samples and relied on evolutionary models \citep[see][and references therein]{ds09a,am12a,tbe11,mm12}. In this paper, we present a new method to derive stellar parameters in a self-consistent manner, combining seismic determinations with the InfraRed Flux Method (IRFM). We compare our results with {\it Hipparcos} parallaxes, high-resolution spectroscopic temperature determinations, and interferometric measurements of angular diameters. We briefly discuss the implications for models of Galactic Chemical Evolution and for age determinations of main-sequence stars.
\section{Sample selection and data extraction}\label{data}
From the more than 500 main-sequence and subgiant stars in which \textit{Kepler} detected oscillations in its short-cadence mode \citep{wc11}, we selected the 32 stars that have {\it Hipparcos} parallaxes \citep{fv07}. We discarded targets with parallax uncertainties larger than 20\% as well as multiple systems and incorrect identifications from the Kepler Input Catalogue \citep[KIC;][]{tb11}. This left us with a final sample of 22 stars. We also retrieved multi-band $B_TV_T$ and $JHK_S$ photometry from the {\it Tycho2} \citep{eh00} and Two Micron All Sky Survey \citep[2MASS;][]{ms06} catalogs, respectively. We have not used the $ugriz$ photometry from the KIC since it suffers from zero-point uncertainties \citep{mp12}, while {\it Tycho2} data have been extensively tested and tied to stellar parameters \citep{lc10}.

Using {\it Kepler} data, corrected as described by \citet{rg11}, the global oscillation observables $\nu_{\rm{max}}$ and $\Delta\nu$ were obtained in two ways. Set A was generated for all 22 stars from the power spectra using the pipeline described by \citet{dh09}, while set B was derived from the individual mode frequencies for the 19 stars for which these are published \citep{ta12}. Set B was based on post-survey data only, meaning that at least three-month-long time series have been used in the analysis. The analysis of the three stars belonging exclusively to set A (namely, KIC 5774694, KIC 5939450 and KIC 10513837) was based on survey data, i.e., of one-month-long duration. Set A includes all of set B.

The way in which seismic parameters were extracted from the frequency lists (set B) deserves a brief explanation: $\nu_{\rm{max}}$ was first estimated by fitting a Gaussian function to the envelope of radial ($l=0$) mode amplitudes as a function of frequency. This least-squares fit was weighted by the uncertainties on the observed amplitudes. Using this value of $\nu_{\rm{max}}$, we computed a proxy of $\Delta\nu$, $\Delta\nu^{\rm{proxy}}$, using the scaling relation given by \citet{ds09a}. This proxy was then used to construct a Gaussian centered at $\nu_{\rm{max}}$ with an FWHM of $4\Delta\nu^{\rm{proxy}}$, which served as the statistical weight when performing a least-squares fit to the radial frequencies as a function of radial order $n$ to determine $\Delta\nu$ \citep[see][]{tw11a,tw11b}. The choice of such a wide weighting envelope around $\nu_{\rm{max}}$ ensures that any frequency dependence of $\Delta\nu$ due to acoustic glitches is averaged out \citep[see][]{bm11,tk12}.

The second step in the selection of a final set of asteroseismic input parameters consisted of performing an internal check on the quoted (uncalibrated) uncertainties. The adopted procedure makes use of the results obtained for the stars common to both sets. We started by computing the rms of the relative residuals in $\nu_{\rm{max}}$ and $\Delta\nu$, i.e., (set~B$-$set~A)/set~A and (set~B$-$set~A)/setB, to be subsequently used in the calibration of the quoted uncertainties given in sets A and B, respectively. This value of the rms, which can be regarded as an additional fractional uncertainty, was then added in quadrature to the uncalibrated uncertainties in order to obtain the final (calibrated) uncertainties.

Since our goal is to perform a uniform analysis, we must use a set of seismic parameters obtained from a single method of extraction. Set A has been selected as the final set (after error calibration) because it covers all the stars. Fractional differences between results in sets A~and~B are less than 3\% for $\nu_{\rm{max}}$ and below 0.5\% for $\Delta\nu$, which are within the uncertainties. The procedure used to obtain set~A has been extensively tested for consistency with other methods of extracting seismic parameters \citep[see][and references therein]{gv11}.
\section{Determining stellar parameters}\label{param}
Our technique for estimating stellar parameters relies on an iterative approach that couples asteroseismic analysis with the results of the IRFM. We do this in such a way that the final values of $\teff$, bolometric flux, mass and radius are dependent on one another and internally consistent.
\subsection{The Direct and Grid-based Methods}\label{aster}
To very good approximation, $\Delta\nu$ scales as the square root of the mean density \citep[e.g.,][]{ru86}, while $\nu_\mathrm{max}$ is related to the acoustic cutoff frequency of the atmosphere \citep[e.g.,][]{tb91,kb95,kbe11}. These two quantities follow scaling relations from the accurately known solar parameters \citep[e.g.,][]{sh09,ds09a}, which can be written as 
\begin{equation}\label{eqn:mass} 
\frac{M}{M_\odot} \simeq \left(\frac{\nu_{\mathrm{max}}}{\nu_{\mathrm{max},\odot}}\right)^{3} \left(\frac{\Delta\nu}{\Delta\nu_\odot}\right)^{-4}\left(\frac{T_\mathrm{eff}}{T_{\mathrm{eff},\odot}}\right)^{3/2}, 
\end{equation}
\begin{equation}\label{eqn:rad} 
\frac{R}{R_\odot} \simeq \left(\frac{\nu_{\mathrm{max}}}{\nu_{\mathrm{max},\odot}}\right) \left(\frac{\Delta\nu}{\Delta\nu_\odot}\right)^{-2}\left(\frac{T_\mathrm{eff}}{T_{\mathrm{eff},\odot}}\right)^{1/2}. 
\end{equation}
Here, $T_{\mathrm{eff},\odot}= 5777\,\rm K$, $\Delta\nu_\odot = 135.1\pm0.1\,\rm \mu Hz$ and $\nu_{\mathrm{max},\odot}= 3090\pm30\,\rm \mu Hz$ are the observed values in the Sun, derived using the same method as set~A \citep{dh11}. Provided a value of $\teff$ is available, these scaling relations give a determination of stellar mass and radius for each star that is independent of evolutionary models \citep[see, e.g.,][]{am09,sh11b,vsa11a}, in what has come to be known as the \textit{direct method}.

Another approach is to include models of stellar evolution when estimating the masses and radii. This so-called \textit{grid-based method} uses evolutionary tracks constructed with a range of metallicities and searches for a best-fitting model, using $\Delta\nu$, $\nu_\mathrm{max}$, $\teff$, and $\feh$ as input parameters \citep[e.g.,][]{ds09b,sb10,ng11,sb12}.

Our reference grid of stellar models was computed with the Garching Stellar Evolution Code \citep[GARSTEC,][]{ws08}. We have used Irwin's equation of state \citep{sc03}, OPAL opacities \citep{ir96} that were complemented at low temperatures by those of \citet{jf05}, and nuclear reaction rates by \citet{ea11}. Models for masses below 1.4~$\msun$ included microscopic diffusion of helium and metals, following \citet{at94}. This effect was not considered for masses above 1.4~$\msun$ since its overall evolutionary impact is small and the code does not include other processes that might become relevant in these cases, such as radiative levitation \citep[e.g.][]{st98}. Core and envelope convective overshooting has been included in all models, using the exponential decay description of \citet{bf96} with a scale factor $f= 0.02$ and a geometric restriction for small convective cores \citep{zm10}. A mixing length parameter of $\alpha=1.811$ from a calibrated solar model and an Eddington $T-\tau$ relation for stellar atmospheres have been adopted.

The grid spans from 0.8 to 2.0~$\msun$ in steps of 0.01~$\msun$. For the metallicity, we defined $\feh= 0$ at the adopted present-day solar photospheric value of $Z/X= 0.0230$ \citep{gs98}. The initial composition of models was, however, defined in terms of that of a calibrated solar model with $Z_0= 0.01876$ and $Y_0=0.26896$. Because of gravitational settling, this implies that the change in $\feh$ of a solar-metallicity track in 4.57~Gyr of evolution is actually $\sim+0.06$~dex. Initial $\feh$ values in the grid range from --0.54 to +0.36~dex in steps of 0.1~dex; the adopted $\Delta\,Y/\Delta\,Z=~1.4$ relation \citep[e.g.,][]{lc07} allows for the complete determination of initial composition of models. The grid was restricted to models with ages above 0.05~Gyr to avoid degeneracy with pre-main-sequence models. The largest time step in the main sequence was constrained to 10~Myr, resulting in a very dense grid that comprises, in total, about $4.5\times 10^6$ models.

When applying the grid-based approach, we obtained the $\Delta\nu$ value of each model using frequencies of individual radial modes calculated with ADIPLS \citep{jcd08}, in the manner described by \citet{tw11b}. On the other hand, $\nu_\mathrm{max}$ was always computed from the acoustic cutoff frequency relation. In order to find the stellar parameters best fitting the input data, we followed a procedure similar to that presented by \citet{sb10}. Briefly, we produced 10,000 Monte~Carlo realizations of sets of input parameters using random Gaussian noise around the central observed values, and calculated the likelihood for models within 3$\sigma$ of all the observables. Considering only those results with a likelihood larger than 95\% of the maximum likelihood, we formed the probability distribution and assigned the uncertainties to be 34\% of either side of the median value.

As it has been extensively discussed in \citet{ng11} and \citet{sb12}, it is important to consider the dispersion in the grid results arising from the use of different evolutionary codes and input physics. For instance, the change in $\teff$ during the main-sequence phase due to microscopic diffusion for stars more massive than $\sim$1.3~$\msun$ can reach approximately 150~K \citep{st98}. Since this effect is not taken into account in our GARSTEC reference grid for masses above $\sim$1.4~$\msun$, when applying the grid-based method we also obtained stellar parameters using the Yale-Yonsei isochrones presented in \citet{sb10}, the evolutionary tracks from Dotter \citep{ad08} and Marigo \citep{lg00,pm08}, the YREC set of models presented in \citet{ng11}, and evolutionary tracks constructed with non-solar values of the mixing length parameter of convection \citep[the "MLT" set from][]{sb12}. These sets of evolutionary tracks were constructed using different microphysics (e.g., equation of state, nuclear reactions, and opacities), as well as assumptions in the metallicity scale and treatment of convection, overshooting and gravitational settling. The final results obtained from the grid-based method encompass these uncertainties using an error calibration process described in Section~\ref{iter}.
\subsection{The InfraRed Flux Method}\label{irfm}
The IRFM is arguably one of the most direct and least model-dependent techniques to determine effective temperatures in stars. It was originally devised to obtain stellar angular diameters with an accuracy of a few percent \citep{bs77,b79,b80}. Our analysis is based on the IRFM described by \citet{lc06,lc10}.

The basic idea is to recover for each star its bolometric $\mathcal{F}_{\rm{Bol}}\rm{(Earth)}$ and infrared monochromatic flux $\mathcal{F}_{\lambda_{\rm{IR}}}$, both measured at the top of Earth's atmosphere. One must then compare their ratio to that obtained from the same quantities defined on a surface element of the star, i.e.~the bolometric flux $\sigma \teff^4$ and the theoretical surface infrared monochromatic flux. For stars hotter than $\sim4200$~K the latter quantity is relatively easy to determine because the near infrared region is largely dominated by the continuum and depends linearly on $\teff$ (Rayleigh-Jeans regime), thus minimizing any dependence on model atmospheres. The problem is therefore reduced to a proper derivation of stellar fluxes, which can then be rearranged to return the effective temperature. Once $\mathcal{F}_{\rm{Bol}}\rm{(Earth)}$ and $\teff$ are both known, the limb-darkened angular diameter, $\theta$, is trivially obtained.

In the adopted implementation, the bolometric flux was recovered using multi-band photometry ({\it Tycho2} $B_TV_T$ and 2MASS $JHK_S$), and the flux outside of these bands (i.e.,~the bolometric correction) was estimated using a theoretical model flux at a given $\teff$, $\feh$ and $\logg$. The infrared monochromatic flux was derived from 2MASS $JHK_S$ magnitudes only. We used an iterative procedure in $\teff$ to cope with the mildly model-dependent nature of the bolometric correction and surface infrared monochromatic flux. For each star, we used the \cite{ck04} grid of model fluxes, starting with an initial estimate of its effective temperature and working at a fixed $\feh$ and $\logg$ until convergence in $\teff$ within $1$\,K was reached.

The uncertainties stemming from the adopted $\feh$ and $\logg$ were taken into account in the error estimate, but their importance is secondary at this stage since the IRFM has been shown to depend only loosely on those parameters \citep[see][]{lc06}. This makes the technique superior to most spectroscopic methods for determining $\teff$ --provided reddening is known-- since the effects of $\teff$, $\logg$ and $\feh$ on the latter are usually strongly coupled and the model dependence is much more important. The metallicity adopted for each star and the coupling of the IRFM with asteroseismic gravities will be discussed in the next section, together with reddening effects.
\subsection{Iterations and Error Determination}\label{iter}
\begin{table*}[!ht]\footnotesize
\centering
\caption{Input parameters for the 22-star Sample. See text for details.}\label{tab:input}
\begin{tabular}{c c c c c c}
\hline\hline
\noalign{\smallskip}
KIC ID & HIP & $\nu_\mathrm{max}$ ($\mu$Hz) & $\Delta\nu$ ($\mu$Hz) & $\feh$ & $E(B-V)$ \\
\noalign{\smallskip}
\hline
\noalign{\smallskip}
 3632418&94112&1144 $\pm$ 31&60.8 $\pm$ 0.2&-0.01&0.024\\
\noalign{\smallskip}
 3733735&94071&2145 $\pm$ 61&92.3 $\pm$ 0.3&-0.10&0.025\\
\noalign{\smallskip}
 4914923&94734&1887 $\pm$ 181&88.7 $\pm$ 0.3&0.17&0.018\\
\noalign{\smallskip}
 5371516&96528&1018 $\pm$ 33&55.4 $\pm$ 0.2&0.13&0.020\\
\noalign{\smallskip}
 5774694&93657&3442 $\pm$ 274&140.2 $\pm$ 4.0&0.01&0.000\\
\noalign{\smallskip}
 5939450&92771&605 $\pm$ 25&30.5 $\pm$ 2.4&-0.01&0.020\\
\noalign{\smallskip}
 6106415&93427&2219 $\pm$ 60&104.3 $\pm$ 0.3&-0.06&0.000\\
\noalign{\smallskip}
 6225718&97527&2338 $\pm$ 66&105.8 $\pm$ 0.3&-0.15&0.010\\
\noalign{\smallskip}
 7747078&94918&946 $\pm$ 26&54.0 $\pm$ 0.2&-0.26&0.018\\
\noalign{\smallskip}
 7940546&92615&1081 $\pm$ 34&58.9 $\pm$ 0.2&-0.04&0.010\\
\noalign{\smallskip}
 8006161&91949&3570 $\pm$ 96&149.3 $\pm$ 0.4&0.34&0.000\\
\noalign{\smallskip}
 8228742&95098&1175 $\pm$ 34&62.1 $\pm$ 0.2&-0.14&0.025\\
\noalign{\smallskip}
 8751420&95362&571 $\pm$ 15&34.6 $\pm$ 0.1&-0.20&0.010\\
\noalign{\smallskip}
 9139151&92961&2695 $\pm$ 74&117.3 $\pm$ 0.3&0.15&0.012\\
\noalign{\smallskip}
 9139163&92962&1685 $\pm$ 45&81.1 $\pm$ 0.2&0.15&0.012\\
\noalign{\smallskip}
 9206432&93607&1859 $\pm$ 50&84.7 $\pm$ 0.3&0.23&0.013\\
\noalign{\smallskip}
10068307&94675&976 $\pm$ 35&54.0 $\pm$ 0.2&-0.13&0.014\\
\noalign{\smallskip}
10162436&97992&1016 $\pm$ 28&55.8 $\pm$ 0.2&-0.08&0.023\\
\noalign{\smallskip}
10454113&92983&2310 $\pm$ 68&105.1 $\pm$ 0.3&-0.06&0.011\\
\noalign{\smallskip}
10513837&91841&191 $\pm$ 7&14.6 $\pm$ 0.2&0.15&0.026\\
\noalign{\smallskip}
11253226&97071&1669 $\pm$ 45&77.0 $\pm$ 0.2&-0.03&0.011\\
\noalign{\smallskip}
12258514&95568&1499 $\pm$ 40&75.0 $\pm$ 0.2&0.13&0.015\\
\noalign{\smallskip}
\hline
\end{tabular}
\end{table*}
As described in Section~\ref{aster}, the asteroseismic methods provide a mass and radius based on an input $\teff$ value (and $\feh$ for the grid-based case). On the other hand, the IRFM gives $\teff$ and the bolometric flux at a given input $\logg$ and $\feh$. In order to determine a unique set of stellar parameters for each star, we iterated the two methods in a consistent way, using both the direct and grid-based approach. A simplified version of this technique was first introduced by \citet{vsa11a}.

We started by calculating sets of IRFM effective temperatures for each star at fixed $\logg=2.0--5.0$ in steps of $0.5~{\rm dex}$; this translates into $\teff$ changes of less than 1\% for each $\logg$ step. The metallicity of the targets must be given as an input, and we have considered them in the following order of preference, according to availability: the latest revision of the Geneva-Copenhagen Survey \citep[GCS;][]{lc11}, spectroscopic determinations from \citet{hb12}, or the value given in the KIC increased by $0.18~{\rm dex}$. The latter is the offset found between GCS and the KIC for the 11 stars common in our sample, and is similar to the $+0.21~{\rm dex}$ offset found by \citet{hb12}.

Reddening must also be specified, and our calculations were made assuming distance dependent extinction values from \citet{rd03}. These were obtained after an iteration in distance as described by \citet{am12b}. If $E(B-V)<0.01$ or no estimate was available we assigned $E(B-V)=0.0$. We list in Table~\ref{tab:input} the input parameters used in our analysis.

The procedure applying the direct method works as follows. Using $\logg$ determinations from the KIC as an initial guess, we interpolated in gravity and computed $\teff$ from the IRFM results. This $\teff$ value, together with $\nu_\mathrm{max}$ and $\Delta\nu$, was fed to the scaling relations to obtain a mass, radius, and thus $\logg$. Interpolating again in gravity gave an updated value of $\teff$, and the procedure was repeated until convergence in $\logg$ and $\teff$ was reached.

We obtained 1$\sigma$ uncertainties of the parameters during the iterations. Uncertainties in the seismic observables were taken into account, as well as variations in the $\teff$ determinations arising from different photometric filters and $\logg$ determinations. The results are affected by the assumed value of extinction, and are mildly dependent on the metallicity considered. To account for possible errors in reddening and composition, we have also computed sets of results at $\log\,g=3.5$, one increasing $E(B-V)$ by $+0.01$ (the decreasing case is essentially symmetric), and another one changing the metallicity by $\pm 0.1$~dex. Moreover, a Monte-Carlo simulation was run to estimate the uncertainties in $\teff$ from random photometric errors. Finally, we added an extra 20~K to the error budget to account for the uncertainty in the zero-point of the temperature scale.

The analysis was repeated using the grids mentioned in Section~\ref{aster} to determine mass, radius, and $\logg$ values at each iteration. In all cases we used as input values the seismic observables and metallicities described above, considering an uncertainty in composition of $\pm 0.1$~dex consistent with what was applied for the IRFM. The final set of stellar parameters from the grid-based method and their corresponding uncertainties were obtained in the same manner as the seismic input parameters (cf.~Section~\ref{data}): we adopted the GARSTEC grid as the reference and performed an error calibration by computing the rms of the relative residuals in mass, radius, and gravity, and adding them in quadrature to the original GARSTEC uncertainties.
\section{Results}\label{res}
The procedure outlined in Section~\ref{iter} provided final values of $\teff$, mass, radius, $\log\,g$, and, as mentioned in Section~\ref{irfm}, the associated bolometric flux and $\theta$. Their corresponding 1$\sigma$ uncertainties were also obtained during the iterations. It is important to mention that the $\logg$ values determined via the direct and grid-based method agree better than 0.03~dex, implying that their $\teff$ values are also in agreement within $3$\,K. Using the asteroseismic radius and $\theta$, it is straightforward to estimate the distance:
\begin{equation}
\centering
d_\mathrm{seis}=C\,\frac{2 R}{\theta}\, ,\label{eq:ang}
\end{equation}
where $C$ is the conversion factor to parsecs. In this manner, we determined asteroseismic distances for our 22 sample targets.

\begin{figure*}[!ht]
\centering
\includegraphics{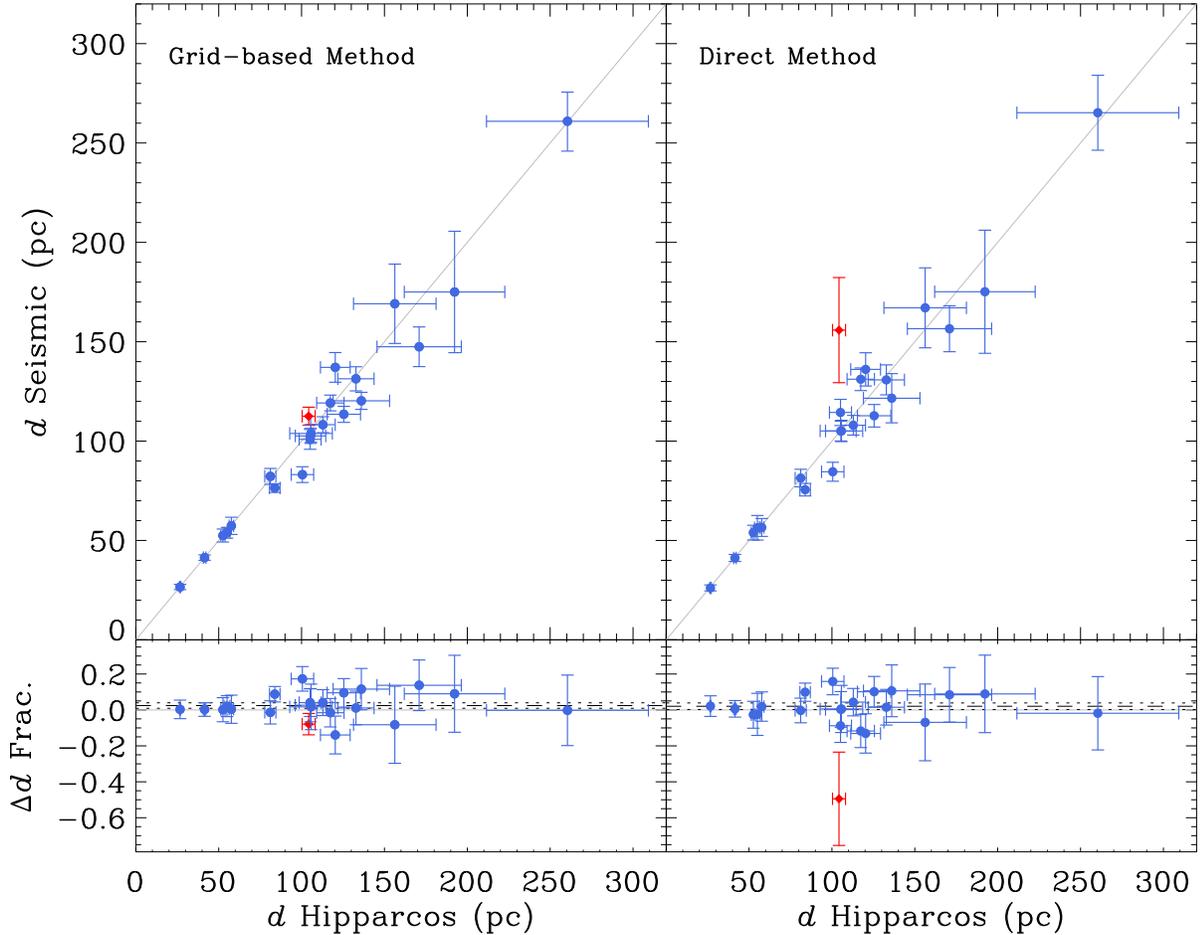}
\caption{Upper panels: comparison of {\it Hipparcos} distances with those obtained via the seismic method. Lower panels: fractional difference $({\rm Hipp.}-{\rm Seis.})$ between both determinations. Gray solid line shows the one-to-one correspondence, while the black dashed and dotted lines represent the weighted average difference and standard deviation, respectively. The red diamond in all panels is the star with the largest fractional error in $\Delta\nu$. See the text for details.}
\label{fig:disGr}
\end{figure*}

In Figure~\ref{fig:disGr} we compare our distances with those obtained from {\it Hipparcos} parallax measurements. Note that, as described in Section~\ref{aster}, seismic radii determinations can be obtained by either the direct or grid-based method. The agreement is excellent, particularly for the close-by targets, boosting our confidence on the asteroseismic parameters and the robustness of our technique.

There is one target that clearly deviates from the one-to-one relation in the results obtained via the direct method, shown with a red diamond in Figure~\ref{fig:disGr}. Not surprisingly, this star has the largest fractional error in $\Delta\nu$ ($\sim$8\%, compared to the less than $\sim$2\% of the rest of the targets, see Equation~\ref{eqn:rad}). The large uncertainty in this parameter dominates the radius error budget, and thus propagates to the estimated distance. For this particular case, the grid-method determined the radius of the star with a much higher accuracy, a result that is confirmed by the better agreement of its distance to that obtained from parallaxes. The fact that the grid-based approach restricts combinations of $\Delta\nu$, $\nu_\mathrm{max}$ and $\feh$ to a narrow range of possible $\teff$ values seems to be sufficient to deal with large uncertainties in one measurement, provided the others are accurately determined \citep{ng11}.

For stars with accurate seismic measurements, the direct method provides results as reliable as the grid-based one. The weighted mean difference $({\rm Hipp.}-{\rm Seis.})$ is $2.1\%\pm1.8\%$ for the direct method, while for the grid-based case is $2.4\%\pm1.5\%$. Removing the outlier from the sample changes the average differences to $2.3\%\pm1.8\%$ and $3.1\%\pm1.6\%$, respectively. One important factor to take into consideration is extinction. From Figure~\ref{fig:disGr} we see that the uncertainties in asteroseismic distances seem to increase with distance. This points out to reddening as the cause, since the error in the seismic distance determinations should be comparable for star with similar uncertainties in the global seismic parameters. Analysis of the error budget shows that reddening becomes the major contributor as distance increases, most likely due to the use of distance-dependent integrated maps of extinction. We discuss this further in Section~\ref{conc}.

As described in Section~\ref{data}, the seismic input parameters were determined using two different methods, and the results shown in Figure~\ref{fig:disGr} are those from set~A (pipeline processing of the power spectrum). We have tested the impact on the distance determinations of using instead the results from individual frequency lists (set~B): for the 19 targets common to both sets, containing stars up to $\sim$170~pc away, the weighted mean difference between the distances of Set~A and Set~B is below 0.5\%.

\begin{figure*}[!ht]
\centering
\includegraphics{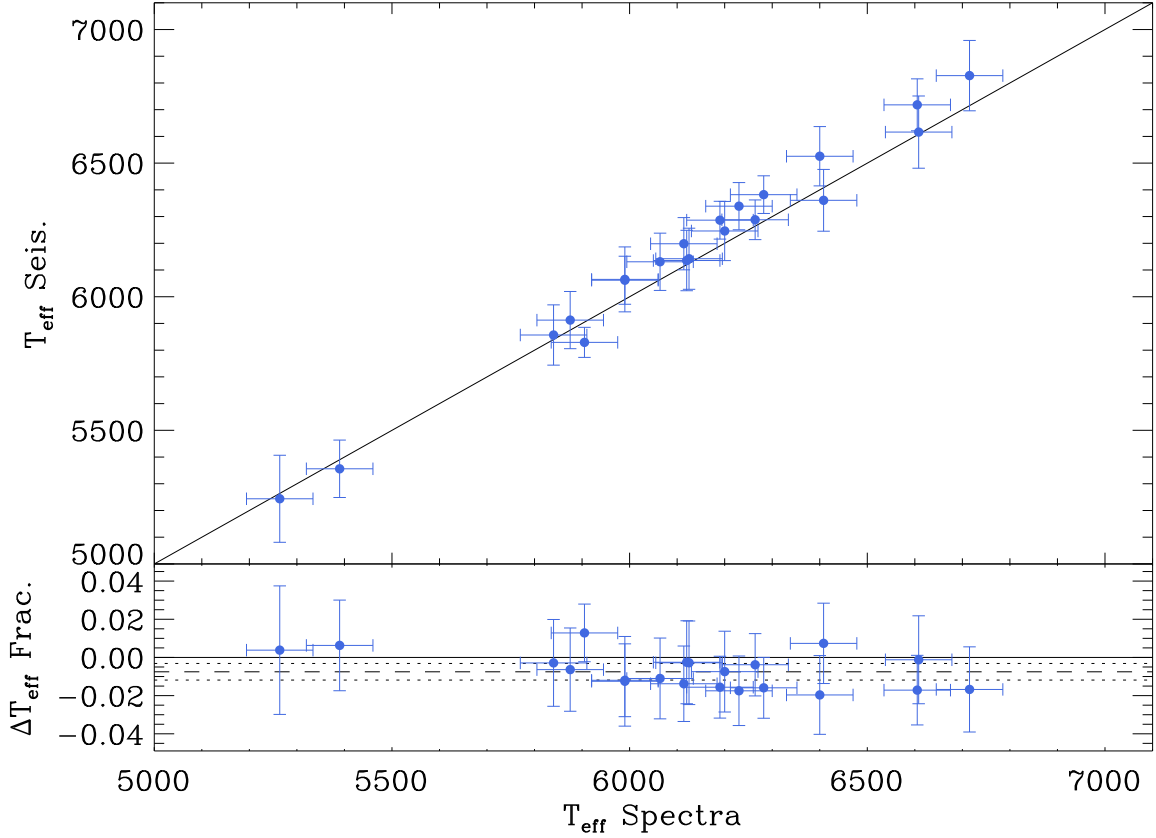}
\caption{Upper panel: comparison of effective temperatures using the asteroseismic method with those obtained from spectroscopy by \citet{hb12}. Lower panel: fractional difference $({\rm Spec.}-{\rm Seis.})$  between both determinations. Solid lines shows the one-to-one correspondence, while the dashed and dotted lines represent the weighted average difference and standard deviation, respectively.}
\label{fig:teff_comp}
\end{figure*}

All but one of our targets were included in the spectroscopic analysis made by \citet{hb12}. Those authors obtained effective temperatures via the excitation balance of Fe~I lines, using a fixed $\logg$ value in their analysis as determined by asteroseismology. In Figure~\ref{fig:teff_comp} we compare their $\teff$ values with ours and find excellent agreement.
Individual fractional differences are all below 2\%, while the weighted mean difference $({\rm Spec.}-{\rm Seis.})$ is $-0.8\%\pm0.4\%$. This level of agreement is particularly impressive considering that the uncertainties quoted by \citet{hb12} are of 70~K for all the targets. However, there is a possible systematic offset to lower values in the spectroscopic temperatures compared to ours. Although the reason for this behavior is far from evident and goes beyond the scope of this paper, we mention that the effective temperatures determined in this work are supported by the analysis of hydrogen line profiles (M.~Bergemann, private communication). The Balmer-line $\teff$ scale is known to be warmer than that given by the excitation balance of Fe~I lines \citep[see][Figure~12]{lc10}, making it more consistent with the IRFM (M.~Bergemann et al. 2012, in preparation).

Another verification of our technique comes from comparing our derived angular diameters with results from interferometry. Four stars in our sample have been observed with the PAVO/CHARA long-baseline interferometer, as described by \citet{dh12}. The residual mean value between our angular diameters and the interferometric ones is below 1\%, consistent with the value found by \citet{dh12} for a larger sample. These results are also compared to the ones derived using the surface brightness technique of \citet{pk04}, obtaining an equally good level of agreement and further confirming the robustness of our method \citep[see Figure~4 in][]{dh12}.

The results outlined above clearly show that our method provides accurate $\teff$, radii, angular diameters, and therefore distances for the sample of stars considered. In order to assess the level of accuracy of these asteroseismic distances, we separated the sample into three bins according to the uncertainty in the parallax measurements. There is a natural correlation between distance and quality of the parallaxes, where stars with smaller uncertainties are usually closer and therefore less affected by interstellar extinction. For those stars with parallax determined better than 5\%, the rms between our grid-based distances and the {\it Hipparcos} ones is below 4.7\%. Moreover, when considering the bin of uncertainties between 5 and 10\%, as well as the bin with parallax errors between 10 and 20\%, the rms of the relative residuals is also smaller than the typical {\it Hipparcos} error. Thus, provided extinction is properly taken into account, our distance determinations can be considered accurate to 5\%.

We present in Table~\ref{tab:stars} the parameters obtained with the grid-based method. Note that, as mentioned in Section~\ref{aster}, the grid-based approach returns a distribution probability function for each parameter, and so the uncertainties in the mass, radius, and gravity are asymmetric.
\begin{table*}[!ht]\tiny
\centering
\caption{Stellar parameters derived using the grid-based method for the 22-star sample. Errors on the parameters come from probability distribution functions (see Sects.~\ref{aster} and~\ref{iter} for details)}\label{tab:stars}
\begin{tabular}{c c c c c c c c c}
\hline\hline
\noalign{\smallskip}
KIC ID & HIP & $\teff$ (K) & $M/M_\odot$ & $R/R_\odot$ & $\logg$ & F$_\mathrm{bol}$ ($10^{-8}$ erg $\mathrm{s}^{-1}\,\mathrm{cm}^{-2}$) & $\theta$ (mas) & $d$ (pc) \\
\noalign{\smallskip}
\hline
\noalign{\smallskip}
 3632418&94112&6286 $\pm$ 70&1.396$^{+0.074}_{-0.075}$&1.911$^{+0.025}_{-0.026}$&4.018$^{+0.003}_{-0.004}$&1.402 $\pm$ 0.074&0.164 $\pm$ 0.006&108$\pm$4\\
\noalign{\smallskip}
 3733735&94071&6824 $\pm$ 131&1.454$^{+0.028}_{-0.055}$&1.427$^{+0.019}_{-0.020}$&4.284$^{+0.008}_{-0.006}$&1.251 $\pm$ 0.070&0.132 $\pm$ 0.006&101$\pm$5\\
\noalign{\smallskip}
 4914923&94734&5828 $\pm$ 56&1.174$^{+0.058}_{-0.062}$&1.408$^{+0.022}_{-0.023}$&4.209$^{+0.003}_{-0.004}$&0.456 $\pm$ 0.023&0.109 $\pm$ 0.003&120$\pm$4\\
\noalign{\smallskip}
 5371516&96528&6360 $\pm$ 115&1.468$^{+0.063}_{-0.080}$&2.066$^{+0.022}_{-0.021}$&3.973$^{+0.007}_{-0.010}$&1.167 $\pm$ 0.062&0.146 $\pm$ 0.007&131$\pm$6\\
\noalign{\smallskip}
 5774694&93657&5911 $\pm$ 106&1.079$^{+0.041}_{-0.045}$&1.000$^{+0.015}_{-0.016}$&4.467$^{+0.007}_{-0.007}$&1.213 $\pm$ 0.076&0.173 $\pm$ 0.008&54$\pm$3\\
\noalign{\smallskip}
 5939450&92771&6380 $\pm$ 70&1.625$^{+0.134}_{-0.069}$&2.916$^{+0.062}_{-0.060}$&3.726$^{+0.008}_{-0.009}$&3.207 $\pm$ 0.169&0.241 $\pm$ 0.008&113$\pm$4\\
\noalign{\smallskip}
 6106415&93427&6061 $\pm$ 89&1.110$^{+0.036}_{-0.033}$&1.240$^{+0.018}_{-0.018}$&4.296$^{+0.004}_{-0.004}$&3.491 $\pm$ 0.024&0.279 $\pm$ 0.008&41$\pm$1\\
\noalign{\smallskip}
 6225718&97527&6338 $\pm$ 88&1.209$^{+0.037}_{-0.034}$&1.256$^{+0.014}_{-0.014}$&4.322$^{+0.004}_{-0.004}$&2.664 $\pm$ 0.291&0.223 $\pm$ 0.014&52$\pm$3\\
\noalign{\smallskip}
 7747078&94918&5856 $\pm$ 112&1.135$^{+0.086}_{-0.088}$&1.952$^{+0.039}_{-0.039}$&3.910$^{+0.004}_{-0.004}$&0.452 $\pm$ 0.099&0.107 $\pm$ 0.012&169$\pm$20\\
\noalign{\smallskip}
 7940546&92615&6287 $\pm$ 74&1.380$^{+0.065}_{-0.104}$&1.944$^{+0.024}_{-0.031}$&3.996$^{+0.006}_{-0.012}$&2.913 $\pm$ 0.039&0.236 $\pm$ 0.006&76$\pm$2\\
\noalign{\smallskip}
 8006161&91949&5355 $\pm$ 107&0.959$^{+0.035}_{-0.037}$&0.927$^{+0.014}_{-0.014}$&4.484$^{+0.004}_{-0.004}$&2.879 $\pm$ 0.153&0.324 $\pm$ 0.016&27$\pm$1\\
\noalign{\smallskip}
 8228742&95098&6130 $\pm$ 107&1.308$^{+0.062}_{-0.060}$&1.855$^{+0.027}_{-0.027}$&4.017$^{+0.005}_{-0.004}$&0.457 $\pm$ 0.156&0.099 $\pm$ 0.017&175$\pm$31\\
\noalign{\smallskip}
 8751420&95362&5243 $\pm$ 162&1.285$^{+0.082}_{-0.096}$&2.722$^{+0.048}_{-0.057}$&3.674$^{+0.006}_{-0.005}$&4.898 $\pm$ 0.368&0.441 $\pm$ 0.032&57$\pm$4\\
\noalign{\smallskip}
 9139151&92961&6141 $\pm$ 114&1.218$^{+0.046}_{-0.046}$&1.178$^{+0.018}_{-0.018}$&4.380$^{+0.004}_{-0.004}$&0.527 $\pm$ 0.017&0.106 $\pm$ 0.004&104$\pm$4\\
\noalign{\smallskip}
 9139163&92962&6525 $\pm$ 111&1.405$^{+0.034}_{-0.027}$&1.571$^{+0.010}_{-0.010}$&4.195$^{+0.004}_{-0.004}$&1.225 $\pm$ 0.031&0.142 $\pm$ 0.005&103$\pm$4\\
\noalign{\smallskip}
 9206432&93607&6614 $\pm$ 135&1.482$^{+0.044}_{-0.044}$&1.544$^{+0.015}_{-0.015}$&4.231$^{+0.005}_{-0.005}$&0.605 $\pm$ 0.064&0.097 $\pm$ 0.007&147$\pm$10\\
\noalign{\smallskip}
10068307&94675&6197 $\pm$ 97&1.366$^{+0.062}_{-0.071}$&2.060$^{+0.028}_{-0.033}$&3.943$^{+0.003}_{-0.004}$&1.401 $\pm$ 0.018&0.169 $\pm$ 0.005&114$\pm$4\\
\noalign{\smallskip}
10162436&97992&6245 $\pm$ 110&1.365$^{+0.061}_{-0.068}$&2.015$^{+0.025}_{-0.027}$&3.961$^{+0.006}_{-0.004}$&0.947 $\pm$ 0.075&0.137 $\pm$ 0.007&137$\pm$7\\
\noalign{\smallskip}
10454113&92983&6134 $\pm$ 113&1.165$^{+0.045}_{-0.045}$&1.251$^{+0.017}_{-0.017}$&4.309$^{+0.005}_{-0.004}$&0.924 $\pm$ 0.050&0.140 $\pm$ 0.006&83$\pm$4\\
\noalign{\smallskip}
10513837&91841&4955 $\pm$ 95&1.290$^{+0.072}_{-0.076}$&4.788$^{+0.083}_{-0.103}$&3.186$^{+0.006}_{-0.007}$&0.585 $\pm$ 0.043&0.171 $\pm$ 0.009&261$\pm$15\\
\noalign{\smallskip}
11253226&97071&6715 $\pm$ 97&1.458$^{+0.032}_{-0.034}$&1.628$^{+0.017}_{-0.018}$&4.176$^{+0.006}_{-0.004}$&1.095 $\pm$ 0.027&0.127 $\pm$ 0.004&119$\pm$4\\
\noalign{\smallskip}
12258514&95568&6064 $\pm$ 121&1.302$^{+0.078}_{-0.084}$&1.630$^{+0.029}_{-0.031}$&4.127$^{+0.004}_{-0.005}$&1.532 $\pm$ 0.063&0.184 $\pm$ 0.008&82$\pm$4\\
\noalign{\smallskip}
\hline
\end{tabular}
\end{table*}

\begin{figure*}[!ht]
\centering
\includegraphics{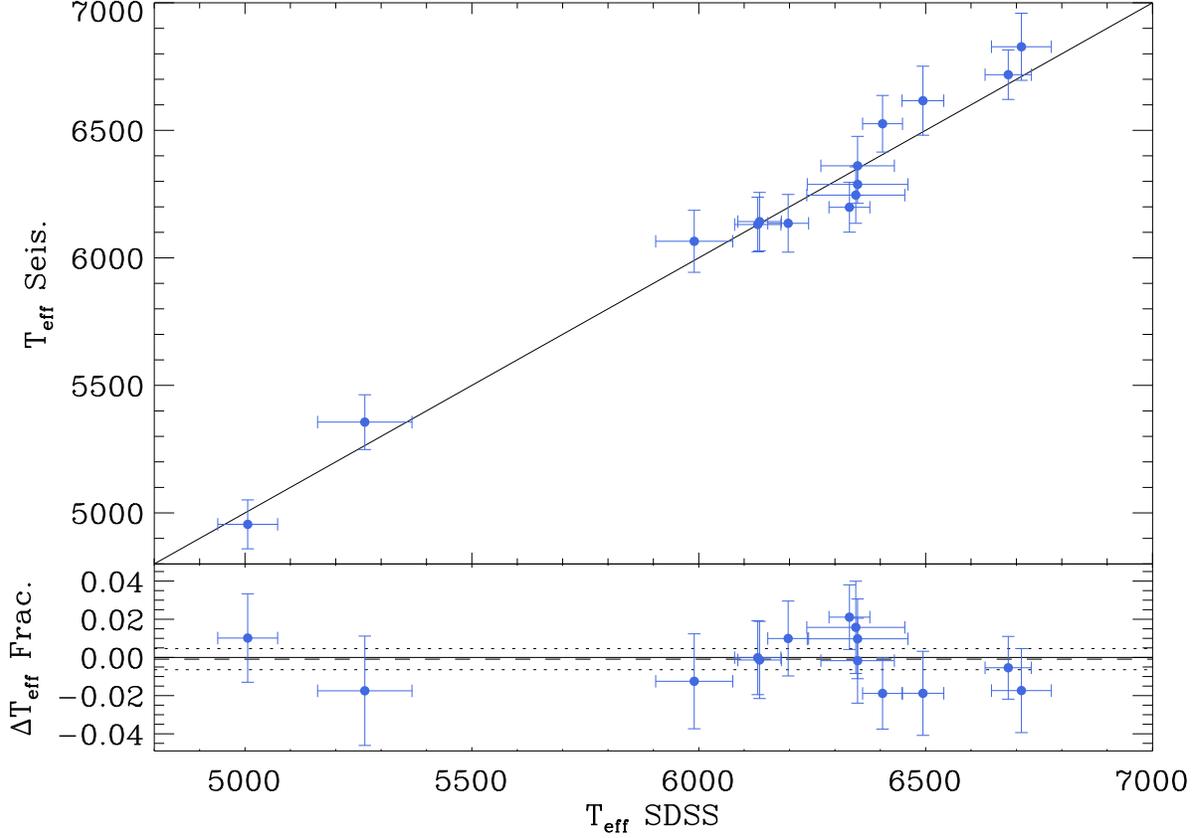}
\caption{Upper panel: comparison of effective temperatures using the asteroseismic method with those derived by \citet{mp12}. Lower panel: fractional difference $({\rm SDSS}-{\rm Seis.})$  between both determinations. Solid lines shows the one-to-one correspondence, while the dashed and dotted lines represent the weighted average difference and standard deviation, respectively.}
\label{fig:teff_comp2}
\end{figure*}

Our parameters can be compared to other studies where the same stellar properties were determined. Recently, \citet{mp12} provided color-temperature relations consistent with the \citet{lc10} scale using the available Sloan Digital Sky Survey (SDSS) $griz$ photometry from the KIC. Fourteen stars from our sample are present in their catalog, and comparison of the obtained $\teff$ values is shown in Figure\ref{fig:teff_comp2}. The weighted mean difference $({\rm SDSS}-{\rm Seis.})$ is $-0.1\%\pm0.6\%$, indicating that the corrected temperatures provided by \citet{mp12} are indeed on a scale consistent with ours and can be used for studies of field stars when this photometry is available.

\citet{sm12} performed a detailed modeling of several {\it Kepler} targets, including six stars from our sample, using individual frequency lists obtained from one--month long observations. The input metallicities and effective temperatures used in that study were in some cases different from ours \citep[see Table~3 in][]{sm12}. Comparing the results reveals good overall agreement in the obtained radii, with two targets showing what appears to be a slight radius underestimation in their determinations with respect to ours. A more thorough investigation of this issue will be made when detailed modeling using longer time series is performed (T.~M. Metcalfe et al. 2012, in preparation).

\begin{figure*}[!ht]   
\centering
\includegraphics{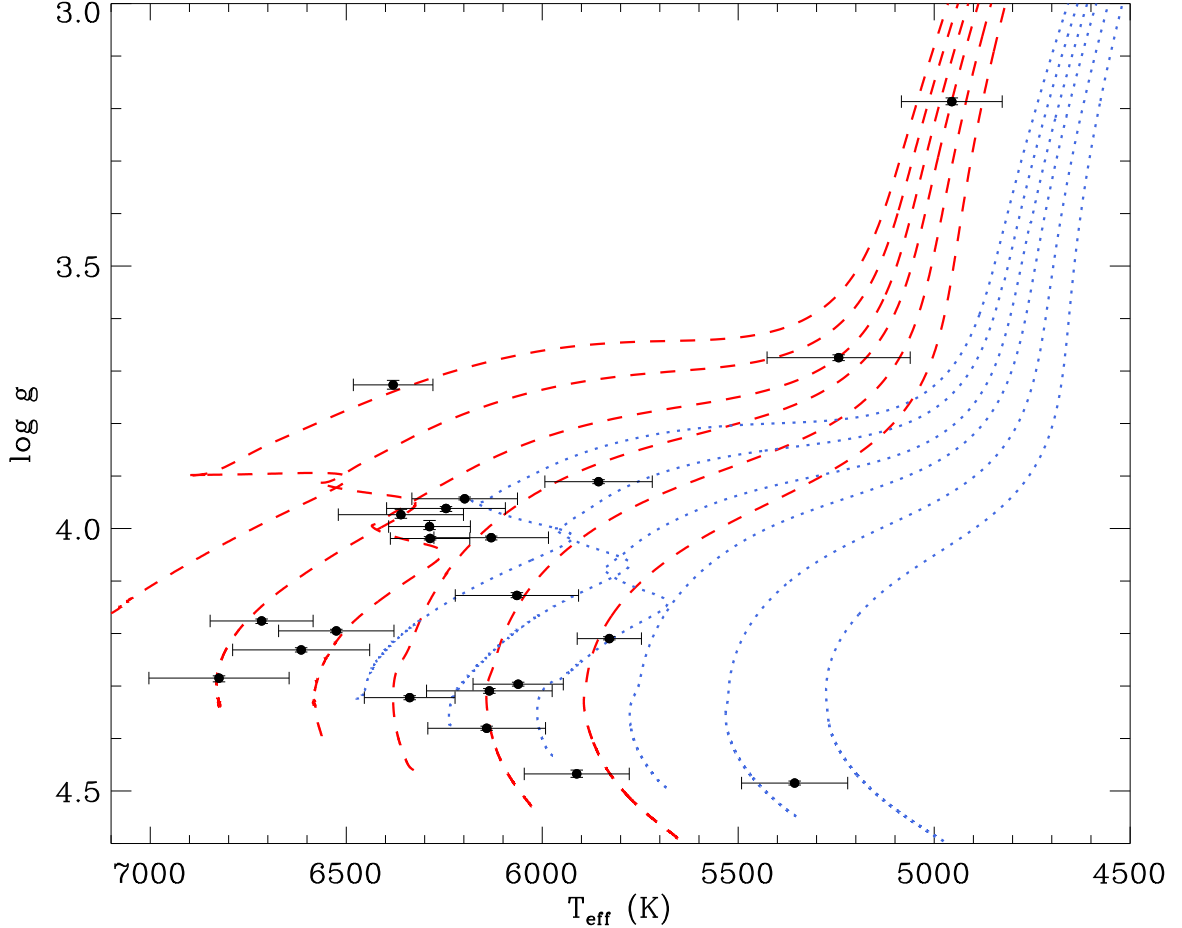}
\caption{Position in the $\logg-\teff$ plane of the 22-star sample as derived via grid-based approach and the IRFM. Also plotted are stellar evolutionary tracks from the GARSTEC grids for masses between 0.9-1.4~$M_\sun$ at two different initial metallicities: $\feh=+0.35$ (dotted blue lines) and $\feh=-0.24$ (dashed red lines).}
\label{fig:hrd}
\end{figure*}

Although our sample of stars only represents a small fraction of the total short-cadence {\it Kepler} sample, they cover a wide range in metallicity, $\teff$ and $\logg$. In Figure~\ref{fig:hrd} we present a $\logg-\teff$ diagram, where the 22~targets have been placed using the parameters derived with the grid-based method. Also plotted are stellar evolution tracks from the GARSTEC grid, at masses and metallicities compatible with those given in Tables~\ref{tab:input}~and~\ref{tab:stars}.
Our targets are distributed in different evolutionary stages, from the early main sequence to the beginning of the red giant branch. Thus, we have tested the accuracy of our method in stars showing a wide variety of masses, ages, compositions, and energy transport mechanisms.

We find good agreement in the radii and bolometric fluxes (and hence distances) estimated by the direct and grid-based methods to the parallax data, which again supports our previous statement that the method has only a modest sensitivity to the stellar composition (see Section~\ref{irfm}). Most importantly, it shows that we can extend the implementation of our method to stars evolved beyond the subgiant branch. This will be addressed in an upcoming publication.

A particularly interesting case is that of KIC~10513837, the most evolved and most distant star in the sample. We estimated its composition to be $\feh=0.15$ by adding 0.18~dex to the KIC value (see Section~\ref{iter} and Table.~\ref{tab:input}). However, its position in Figure~\ref{fig:hrd} is instead compatible with sub-solar metallicity. Spectroscopic analysis of this target made by \citet{jm08} found a value of $\feh=-0.07$ and an effective temperature consistent with our determination, further confirming the mild sensitivity of our method to composition.

\begin{figure*}[!ht]   
\centering
\includegraphics{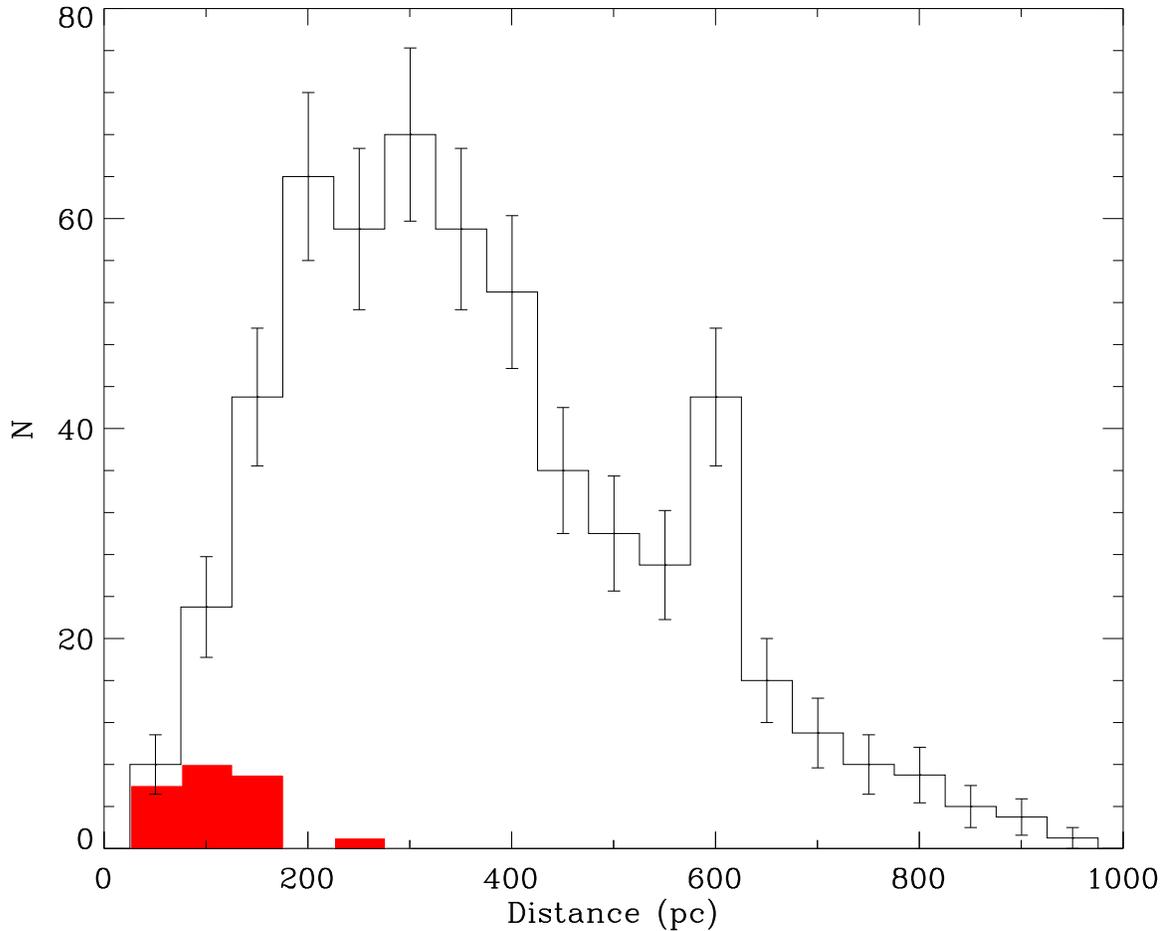}
\caption{Derived distance distributions using ${\rm IRFM}+{\rm asteroseismology}$. Red shaded region shows the results for the 22-sample stars, while the unshaded one depicts the results for all main-sequence and subgiant stars with available \textit{Kepler} asteroseismic data.}
\label{fig:histo}
\end{figure*}

In a natural continuation of this work, we applied our procedure using the direct method to the complete short-cadence \textit{Kepler} sample and derived consistent parameters, including distances, for all these stars. The 565 targets considered are predominantly main-sequence and subgiant stars, with a handful of red giants also present in the sample. All the targets have {\it Tycho2} photometry available, and we have used metallicites from the KIC increased by $0.18~{\rm dex}$. To account for the uncertainties in composition, we have computed IRFM sets of $\teff$ varying the metallicity by $\pm 0.3$~dex (see Section~\ref{iter}). The seismic input parameters were computed as for set~A, described in Section~\ref{data}. Extinction values were obtained from \citet{rd03} after an iteration in distance. These preliminary determinations will be compared with results from several pipelines in a forthcoming publication (W.~J. Chaplin et al. 2012, in preparation). In Figure~\ref{fig:histo} we show a histogram with the obtained distance distribution, where we have also plotted the distribution of our 22 sample stars for comparison.

The resulting distance distribution shows that we can use \textit{Kepler} data to probe populations of main-sequence and subgiant field stars as far as 1~kpc from the Sun. However, our distance determinations for this sample can be slightly undermined by faulty metallicities from the KIC \citep[see][for a discussion]{jm11,hb11}. We have estimated $\feh$ for these stars as consistently as possible with the available data. The latest revision of the GCS is built upon 1500 stars with spectroscopic measurements agreeing with our $\teff$ scale \citep{lc11}. Therefore, on average the $+0.18~{\rm dex}$ correction we find makes the metallicity scale of the KIC consistent with the underlying $\teff$ scale we are deriving here. On the other hand, the good agreement with the results of \citet{hb12} implies that also the latter metallicity scale is broadly on the same zero point that we adopt (and in fact, a similar +0.21 dex correction is found). Thus, although the most self-consistent approach would be to iterate our $\teff$ also for deriving metallicities, on average this condition is fulfilled.

As a final test of the impact on distances of incorrect metallicities, we have done calculations considering different assumptions for the composition of the full sample: using the $\feh$ values as provided in the KIC (i.e., not including the  $+0.18~{\rm dex}$ offset), and also considering a fixed mean metallicity for all the targets of $\feh=-0.2$~dex. The distance distribution obtained remained practically unchanged, reinforcing the notion of the mild metallicity dependence of our method.
\section{Conclusions}\label{conc}
Determining accurate stellar parameters is crucially important for detailed studies of individual stars, as well as for characterizing stellar populations in the Milky Way. The asteroseismic revolution produced by the {\it CoRoT} and \textit{Kepler} missions requires robust techniques to exploit fully the potential of the data and provide the community with the building blocks for ensemble analysis.

Using oscillation data and multi-band photometry, we have presented a new method to derive stellar parameters, combining the IRFM with asteroseismic analysis. The novelty of our approach is that it allows us to obtain radius, mass, $\teff$, and bolometric flux for individual targets in a self-consistent manner. This naturally results in direct determinations of angular diameters and distances without resorting to parallax information, further enhancing the capabilities of our technique.

Two asteroseismic methods were applied to the available data, one based entirely on scaling relations and the other one using grids of pre-calculated models. When accurate seismic data are available, comparison of our distance results with those from {\it Hipparcos} parallaxes shows an overall agreement better than 4\%, regardless of the asteroseismic method employed. Furthermore, the obtained $\teff$ values show a mean difference below 1\% when compared to results from high-resolution spectroscopy.  We have also compared our calculated angular diameters with those measured by long-baseline interferometry and found agreement within 5\%. This provides verification of our radii, $\teff$ values and bolometric fluxes to an excellent level of accuracy. 

Despite the encouraging results, systematics can arise from faulty determinations of reddening values and metallicities. In Section~\ref{iter} we described how the effects of metallicity and reddening, as well as their uncertainties, were taken into account in the calculations. For our determinations to be completely self-consistent, we must be able to determine those parameters from a single set of data using the IRFM. An observational campaign is currently under way to obtain Str\"omgren photometry of \textit{Kepler} stars that will provide a homogeneous set of values for $\feh$ and extinction, as described by \citet{lc11}.

For most of the stellar parameters included in our verifications, both asteroseismic methods produce equally good results. However, it should be kept in mind that the direct method can be significantly biased when large uncertainties in the seismic input parameters exist. Moreover, scaling relations are likely to have a different dependance on effective temperature beyond the main-sequence phase, as suggested by comparison to evolutionary calculations \citep[see][]{tw11b}. The restrictions imposed by metallicity and by the theory of stellar evolution help to cope better with large errors in seismic data, and the use of the grid-based analysis in these cases is therefore recommended . 

However, to take full advantage of the available parameters, asteroseismology must provide masses with a comparable level of accuracy. It is important to note that results on masses from the direct method for values above $\sim$~1.5~$M_\odot$ can deviate significantly from those obtained using the grid-based approach. In fact, differences of more than $\sim$30\% are not unusual in these cases. Using different grids of models, \citet{ng11} found that the fact that the direct method does not explicitly take metallicity into account could undermine its mass determinations. A thorough comparison of different grid-based techniques with the direct method is beyond the scope of this paper and will be presented in an upcoming publication (W.~J. Chaplin et al. 2012, in preparation). Another method to obtain asteroseismic masses is via detailed modeling of targets, aiming at fitting the list of individual frequencies \citep[e.g.][]{tm10,sm12}. This approach provides mass estimates with a high level of precision and, in principle, also with high accuracy. Regardless of the considered technique, one must keep in mind that verification of asteroseismic mass determinations in general is still needed.

Studies of the stellar populations in the {\it CoRoT} and {\it Kepler} fields can greatly benefit from accurate masses, radii, $\teff$, and distances \citep{am12a,am12c}. Combining this information with evolutionary models can lead to an age-metallicity relation, opening the possibility of testing models of Galactic Chemical Evolution in stars outside the solar neighborhood \citep[e.g.,][]{cc97,sb09,kf12}. Applying our method to the complete short-cadence {\it Kepler} sample reveals that we can probe stars as far as 1~kpc from our Sun, making this set of main-sequence and subgiant stars extremely interesting for population studies. Although much greater distances can be probed by analyzing oscillations in giants, the ages of these stars are mostly determined by their main-sequence lifetime \citep[e.g.,][]{ms02,sb11}. Thus, the short-cadence sample is of key importance for helping to calibrate mass-age relationships of red giants and correctly characterize their populations.

A substantial number of the {\it Kepler} main-sequence and subgiant targets have been observed long enough to obtain individual frequency determinations \citep{ta12}. Detailed modeling of these stars, particularly using frequency combinations\citep[e.g.,][]{rv03,pdm10} and modes of mixed character \citep[e.g.,][]{dm11,ob12}, can put tighter constraints on their masses and ages, providing anchor points for ensemble studies. In fact, certain combinations of frequencies can be used to probe the remaining central hydrogen content in stars \citep[e.g.,][]{jcd88}, the existence and size of a convective core \citep[e.g.,][and references therein]{vsa11b}, and the position of the convective envelope and helium surface abundance \citep[e.g.,][]{jcd91,ab94}. These techniques are currently being applied to several stars in the sample (e.g., S. Deheuvels et al. 2012, in preparation, A. Mazumdar et al. 2012, in preparation, V. Silva Aguirre et al. 2012, in preparation) and should help us obtain masses with higher accuracy and determine more robust differential ages.
\acknowledgements
Funding for this Discovery mission is provided by NASA's Science Mission Directorate. The authors thank the entire \emph{Kepler} team, without whom these results would not be possible. We also thank all funding councils and agencies that have supported the activities of KASC Working Group\,1. We are also grateful for support from the International Space Science Institute (ISSI). The authors acknowledge the KITP staff of UCSB for their warm hospitality during the research program ``Asteroseismology in the Space Age''. This KITP program was supported in part by the National Science Foundation of the United States under Grant No. NSF PHY05Ð51164. V.S.A.\ received financial support from the {\sl Excellence cluster ``Origin and Structure of the Universe''} (Garching). Funding for the Stellar Astrophysics Centre is provided by The Danish National Research Foundation. The research is supported by the ASTERISK project (ASTERoseismic Investigations with SONG and {\it Kepler}) funded by the European Research Council (Grant agreement No. 267864). S.B. acknowledges NSF grant AST-1105930. T.L.C. and M.J.P.F.G.M. acknowledge financial support from project PTDC/CTE-AST/098754/2008 funded by FCT/MCTES, Portugal. W.J.C. and Y.E. acknowledge the financial support of the UK Science Technology and Facilities Council (STFC). D.H. is supported by an appointment to the NASA Postdoctoral Program at Ames Research Center, administered by Oak Ridge Associated Universities through a contract with NASA. A.M.S. is partially supported by the European Union International Reintegration Grant PIRG-GA-2009-247732 and the MICINN grant AYA2011-24704. S.H. acknowledges financial support from the Netherlands Organisation for Scientific Research (NWO). D.S. acknowledges support from the Australian Research Council. NCAR is partially funded by the National Science Foundation.

\end{document}